# Heart Rate Variability as an Indicator of Thermal Comfort State

Kizito N. Nkurikiyeyezu[1†], Yuta Suzuki[1], Yoshito Tobe[1], Guillaume F. Lopez[1], and Kiyoshi Itao[2]

[1]Graduate School of Science and Engineering, Aoyama Gakuin University, Kanagawa, Japan
kizito@wil-aoyama.jp, ysuzuki@wil-aoyama.jp, yoshito-tobe@rcl-aoyama.jp, guillaume@wil-aoyama.jp
[2]The University of Tokyo 7-3-1 Hongo, Bunkyo-ku, Tokyo 113-0033, Japan
itao@npowin.org

**Abstract:** Thermal comfort is a personal assessment of one's satisfaction with the surroundings. Yet, most thermal comfort delivery mechanisms preclude physiological and psychological precursors to thermal comfort. Accordingly, many people feel either cold or hot in an environment that is supposedly thermally comfortable to most people. To address this issue, this paper proposes to use people's heart rate variability (HRV) as an alternative indicator of thermal comfort. Since HRV is linked to homeostasis, we hypothesize that it could be used to predict people's thermal comfort status. To test our hypothesis, we analyzed statistical, spectral, and nonlinear HRV indices of 17 human subjects doing light office work in a cold, a neutral, and a hot environment. The resulting HRV indices were used as inputs to machine learning classification algorithms. We observed that HRV is distinctively altered depending on the thermal environment and that it is possible to steadfastly predict each subject's thermal environment (cold, neutral, and hot) with up to a 93.7% prediction accuracy. The result of this study implies that it could be possible to design automatic real-time thermal comfort controllers based on people's HRV.

## 1. INTRODUCTION

Thermal comfort is largely dispensed using heat-transfer mathematical models. These models, however, only reflect the influence of the environment on the person's thermal comfort but relate neither to the complexity of human thermoregulation nor to the adequacy of the provided thermal comfort. Moreover, they ignore occupants' psychophysics, gender, age and other physiological, psychological, cultural and social contexts that are known to affect the perception of thermal comfort [1]–[4]. Consequently, in practice, they fail to deliver an optimum thermal comfort [5]–[8].

In this paper, we propose to use people's heart rate variability (HRV) to estimate people's state of thermal comfort (cold, comfortable, or hot). HRV is a non-periodic time variation between two consecutive heartbeats (R-R intervals). This variation is, nonetheless, not random; instead, it changes depending on complex interactions between the parasympathetic and the sympathetic nervous system.

Because thermal comfort is, by definition, a personal subjective assessment of the satisfaction of the mind with the thermal environment [9], and given that thermal changes in environment affect homeostasis, which in turn affects HRV [10], we surmised that one's state of thermal comfort could be anticipated based on his HRV.

## 2. MATERIAL AND METHODS

We recorded electrocardiogram (ECG) signals of 17 human subjects doing light office work. The experiment was set up in an artificial cold, neutral, and hot thermal chamber (Table 1). Each experiment lasted for about 30 minutes. To analyzed the recorded ECG, we developed an ad-hoc HRV analysis software that calculated statistical, spectral, and nonlinear HRV indices summarized in Table 2. The extraction of R-R interval signals from raw ECG signals is based on the Hamilton-Tompkins algorithm [11] and the HRV indices are calculated according to standards and algorithms proposed in [12] and [13]. To compute the HRV indices, for each subject, the HRV indices were first extracted from a five-minutes-long ECG window segment. Then, the window segment is shifted by approximately 15 seconds and new HRV indices are calculated. This process is repeated until the end of the entire ECG recording. The computed HRV indices were then used as input to supervised Machine Learning (ML) algorithms to classify the state of thermal comfort (cold, comfortable, and hot) of each subject. Randomized Logistic Regression (RLR) algorithms [14] were used to select the most vital HRV indices for thermal comfort classification.

**TABLE 1.** Climate chamber thermal settings

|  | Cold | Neutral | Hot |
|---|---|---|---|
| Activity level | 1 | 1 | 1 |
| Clothing level | 1 | 1 | 1 |
| Air temperature (℃) | 18.0 | 24.0 | 30.0 |
| Mean radiant temperature (℃) | 18.0 | 24.0 | 30.0 |
| Air speed (m/s) | 0.3 | 0.3 | 0.3 |
| Humidity (%RH) | 50.0 | 65.0 | 80.0 |
| PMV Index | -1.89 | 0.02 | 2.09 |

PMV = Predicted Mean Vote

## 3. RESULTS AND DISCUSSION

In general, subjects reported feeling cool or cold in the cold chamber, comfortable in the neutral environment and warm or hot in a hot environment. However, the level of thermal sensation slightly fluctuated from on subject to another. While there is a possibility that this variation is due to a bias of self-report questionnaires, other studies have shown that different people have different thermal preferences [15]–[19]. This discrepancy is due, in part, to variations in physiological and physiological between people. It may be also a result of differences in thermal perception between subjects. In all cases, however, no subject expressed being hot in a cold environment or being cold in a hot environment; thus, we concluded that the subjects felt cold in a cold environment and hot in a hot environment

A comparison of HRV in the three environments revealed that HRV is distinctly altered from one environment to another.

**TABLE 2** – Summary of the HRV indices

| | Statistical time domain HRV indices |
|---|---|
| Mean RR | Mean of R-R intervals |
| RMSSD | Square root of the mean of the sum of difference of successive R-R intervals |
| SDSD | Standard deviation of difference between adjacent R-R intervals |
| pNNx | Percentage of number of R-R pairs that differ by x milliseconds in the entire recording |
| | Frequency Domain HRV indices |
| TP | Total spectral power (0-0.4Hz) |
| VLF | Power in very low range frequencies (0.003 – 0.04Hz) |
| LF | Power in low range frequencies (0.04 –0.15 Hz) |
| HF | Power in high range frequencies (0.15 – 0.4 Hz) |
| LF/HF | Ration between LF and HF power |
| | Non-linear HRV indices |
| DFA($\alpha_1$) | Short-term fluctuations of the Detrended Fluctuation Analysis (DFA) |
| Samp. En | Sample Entropy – A measure of predictability |

The short-term DFA ($\alpha 1$) was highest in the hot environment and mostly lowest in the cold environment (Fig. 1). Other HRV indices such as the mean RR, the VLF, and the sample entropy were highest in the cold environment and lowest in the hot environment (Fig. 2).

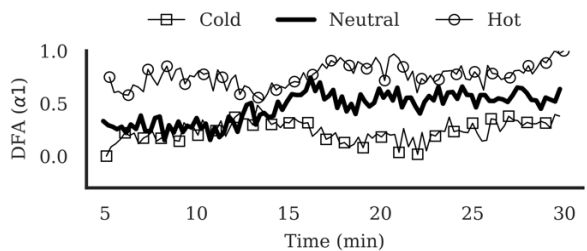

*Fig.1 Normalized mean of all subjects' DFA ($\alpha_1$) indicating that this HRV index is consistently highest in the hot environment and mostly lowest in the cold environment*

Randomized Logistic Regression (RLR) feature selection algorithms [14] showed that the Mean RR, the RMSSD, the SDSD, the pNN25, the VLF and the sample entropy HRV indices are the most important HRV indices for thermal comfort prediction.

These HRV indices were used to predict the thermal environment of each subject given his HRV indices. The highest performance was achieved by a Support Vector Machine (SVM) model at 93.7% accuracy. Nevertheless, even simpler, and less computational intensive machine learning classification models such as the Naive Bayes (NB) achieved a decent 88.5% prediction accuracy.

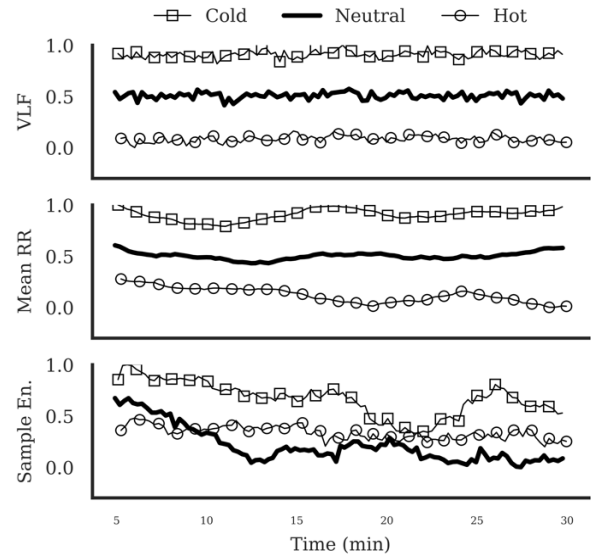

*Fig. 2 Normalized mean of all subjects' VLF, Mean RR and Sample Entropy HRV indices indicating that these HRV indices are consistently highest in the cold environment.*

## 3. CONCLUSION

This paper proposed HRV as an alternative indicator for thermal comfort in office environments. Unlike existing mechanisms, this approach would allow predicting thermal comfort based on people's physiological response to the surrounding thermal environment. Our study found that HRV distinctively varies depending on the thermal environment and that it is feasible to consistently predict each subject's thermal environment (cold, neutral, and hot) with up to a 93.7% accuracy.

Based on these findings, we believe it could be possible to design computerized systems that regulate thermal comfort based on the physiological sensation of office occupants. Further, since previous studies have indicated that most people complain about too hot or too cold temperatures in buildings [20] and thermal neutrality in buildings does not necessarily reflect thermal comfort —since, in reality, most people prefer non-neutral conditions [15]–[18]— we believe that it could be possible to allow indoor office temperatures to drift away from the neutral thermal conditions and readjust the indoor temperature only when the office occupants are feeling either too cool or too warm. This approach is expected to reduce energy consumed for thermal comfort provision [21] and could be achieved without significantly sacrificing people's thermal comfort. Moreover, due to the availability of unobtrusive ECG sensors and a proliferation of inexpensive wearable devices with built-in HRV monitoring capability (e.g. smartwatches and fitness trackers), this ambition could be achieved in a non-invasive manner and at low cost. We are developing computerized systems to achieve this.